\documentclass{revtex4-2}[amsmath,amssymb,aps,floatfix]

\usepackage{graphicx}
\graphicspath{{./figures/}}
\usepackage{dcolumn}
\usepackage{bm}
\usepackage{float}
\usepackage{amsmath}
\usepackage{amssymb}
\usepackage{xcolor}

\definecolor{dgreen}{RGB}{0, 128, 43}

\begin{document}

\title{Nonlinear Shaping in the Picosecond Gap}

\author{Randy Lemons$^1$}
\email{rlemons@slac.stanford.edu}
\author{Jack Hirschman$^{1,2}$}
\author{Hao Zhang$^{1,4}$}
\author{Charles Durfee$^3$}
\author{Sergio Carbajo$^{1,4}$}

\affiliation{$^1$SLAC National Accelerator Laboratory, 2575 Sand Hill Rd, Menlo Park, CA 94025}
\affiliation{$^2$Stanford University, 348 Via Pueblo, Stanford, CA 94305}
\affiliation{$^3$Colorado School of Mines, 1500 Illinois St, Golden, CO 80401}
\affiliation{$^4$University of California, Los Angeles, Los Angeles, CA 90095}

\date{\today}

\begin{abstract}
Lightwave pulse shaping in the picosecond regime has remained unaddressed because it resides beyond the limits of state-of-the-art techniques, either due to its inherently narrow spectral content or fundamental speed limitations in electronic devices. The so-called picosecond shaping gap hampers progress in all areas correlated with time-modulated light-matter interactions, such as photoelectronics, health and medical technologies, and energy and material sciences. We report on a novel nonlinear method to simultaneously frequency-convert and adaptably shape the envelope of light wavepackets in the picosecond regime by balancing spectral engineering and nonlinear conversion in solid-state nonlinear media, without requiring active devices. We capture computationally the versatility of this methodology across a diverse set of nonlinear conversion chains and initial conditions. We also provide experimental evidence of this framework producing picosecond-shaped, ultra-narrowband, near-transform limited light pulses from broadband, femtosecond input pulses, paving the way toward programmable lightwave shaping at GHz-to-THz frequencies.
\end{abstract}

\maketitle

\section{Main}
Controlling the time-domain intensity shape of ultrashort laser pulses is crucial for various applications in science and technology. Advanced intensity-only modulation in the frequency domain allows for the generation of arbitrary waveforms with controlled amplitude and phase distributions opening new possibilities for applications in microscopy, time-resolved imaging, laser micro-machining, particle acceleration, and attosecond science\cite{bejot2022taming, forbes2021structured}.
Managing the temporal characteristics of ultrashort laser pulses has been an active area of development since their inception in the early 1960s. The predominant techniques that have emerged to address this can be broadly classified as either spectral or temporal techniques. Spectral methods specialize in modulating the spectral amplitude and/or phase of broadband optical pulses, typically in the visible to near-infrared (NIR) range using devices such as spatial light modulators\cite{mironov2016shaping,penco2013optimization} or acousto-optic modulators\cite{li2009laser,petrarca2007production} in the Fourier plane of a dispersive element. The main limitation to utilizing spectral methods for pulse shaping is having sufficient bandwidth for the modulation devices to act upon, resulting in a limitation on input pulse durations on the order of 100 fs or less. Temporal techniques directly modulate the temporal amplitude of pulses most commonly using electro-optic modulators\cite{DIELS2006433,SkeldonEOMshaping}. However, the time resolution with direct temporal shaping is limited by the electronic response time of the modulation devices, typically on the order of a nanosecond and above, thus limiting the application of temporal methods to pulses with similar or longer durations. As such, the combined applicable temporal regimes across both methodologies do not include shaping pulses with transform-limited (TL) durations spanning from a few to hundreds of picoseconds due to the bandwidth of these pulses being too narrow for spectral methods and the duration too short for direct temporal methods.

Achieving efficient control of the spatiotemporal profile of pulses in this region has numerous and immediate applications. For example, certain reactions during photosynthesis have characteristic times on the order of tens of picoseconds\cite{guo2024closing} potentially opening the door for artificial photosynthesis driven by shaped optical pulses with tens of picosecond duration\cite{baikie2023photosynthesis}. The study of photoexcitation and charge carrier dynamics in semiconductors with complex dynamics on the picosecond scale can improve optoelectronic devices for faster information processing\cite{perez2024picosecond}. Similarly, new frontiers in ultra-fast picosecond-duration electronics\cite{samizadeh2020nanoplasma} can be opened by patterning of picosecond-duration pulses driving nanoplasma excitation\cite{sun2024all}. Furthermore, X-ray free electron lasers, the brightest X-ray sources available today, are driven by picosecond duration photoexcitation laser where the spatiotemporal profile has a drastic impact on the space-space of photoelectrons and thus on the temporal, spatial, and spectral properties of X-ray pulses \cite{PhysRevSTAB.5.094203,penco2014experimental,zhang2024linac}.

Successful shaping efforts in this picosecond shaping gap have mainly relied upon stretching femtosecond broadband pulses out to picosecond duration and modulating either the spectral amplitude\cite{Agostinelli:79,Haner:87} or spectral phase\cite{zhang2017high,Weiner:88}. Once the broadband pulses are sufficiently dispersed, the temporal amplitude of the pulse can be roughly approximated by the spectral amplitude profile allowing for a linear mapping of the spectral techniques into the temporal domain. Recently, the spectral phase transfer technique for temporal pulse shaping has been successfully demonstrated using four-wave mixing (FWM) in hollow-core fibers~\cite{zhang2024optimizing}. Using the nonlinear optical properties of the fiber enables precise control over the temporal structure of pulses. The desired phase profile can be effectively mapped onto the idler pulse generated through the FWM process by injecting a signal pulse with a predefined spectral phase. This approach opens new possibilities for high-precision temporal shaping applications, particularly under high-power regimes. However, its extension to the picosecond regime remains challenging due to transfer efficiency limitations. The difficulty arises from the complex interplay of nonlinear effects and phase-matching conditions required for efficient spectral phase transfer over shorter pulse durations. 
Alternatively, a shaped pulse can be generated by stacking many copies of few-picosecond duration pulses with increasing delay and varying amplitude\cite{will2008generation}. While these methods are in principle suitable for achieving temporal intensity shaping in the picosecond regime, they exhibit unavoidable downsides in the form of residual spectral chirp or intensity fluctuations due to phase errors. As such, the use of pulses is difficult or impossible in phase-sensitive or intensity-sensitive applications such as nonlinear frequency conversion, micromachining\cite{jiang2024review}, and photoexcitation\cite{PhysRevAccelBeams.23.024401,mitchell2016sensitivity,zhang2024linac} where even 5$\%$ deviations from expected can have drastically different measurable results.

Here we present a broadly applicable and versatile framework with experimental evidence for generating tailored temporal envelopes with (near-)TL spectral bandwidths in the picosecond regime. The baseline operating principle mixes a pair of broadband pulses that have been oppositely chirped with tailored second- and third-order spectral dispersion in a non-collinear sum-frequency generation (SFG) scheme. During the nonlinear interaction, the resultant SFG pulse is generated with a duration commensurate with the stretched duration of the mixed broadband pulses – an intensity envelope that is roughly equal to the sum of the input pulses’ envelope in the time domain – and an ultra-compressed spectral bandwidth down to the picometer. The bandwidth reduction and summing of opposite spectral chirps result in a pulse that has a remarkably low residual spectral phase noise and fluctuations, thus overcoming the limitations of any prior art. We extend beyond this baseline by adding spectral amplitude shaping onto the mixed input pulses to the SFG in addition to spectral phase control.

\section{Results}
\subsection{Theoretical Description}

Our framework, which we have coined as dispersion controlled nonlinear synthesis (DCNS)\cite{lemons2022temporal}, builds on the principle behind spectral compression\cite{raoult1998efficient,ribeyre2001all} wherein an initial TL broadband – e.g., few nm-bandwidth and beyond in the optical range – pulse is split into two identical copies. Here, a significant amount of equal and opposite second-order dispersion (SOD) is added to the spectral phase of each copy, and then the two are mixed in a non-collinear SFG scheme to generate a near-TL picosecond duration pulse. DCNS builds on these principles by adding third-order dispersion (TOD) to the baseline SOD when modifying the spectral phase and spectral amplitude shaping before nonlinear conversion. Unlike SOD, the added per-pulse TOD and spectral filtering are not held to the equal and opposite requirement like SOD. In theory, the SOD could also be adjusted individually, however, the nonlinear conversion efficiency is quickly decreased as the SODs of pulse pair become mismatched\cite{kuzmin2021highly}. By including these two shaping factors for the two incident pulses, we add four additional degrees of freedom for shaping the temporal intensity profile of the SFG pulse drastically increasing the effective parameter space. For each pulse, the applied spectral phase is then a combination of both SOD ($\varphi_2$) and TOD ($\varphi_3$), where $\varphi_2$ is the primary control on pulse duration and $\varphi_3$ controls the final SFG pulse envelope shape. However, if the applied $\varphi_2$ is increased to generate a longer SFG pulse, TOD must also increase to maintain the desired shape. This relationship is linear and by defining the ratio between TOD and SOD, $\alpha = \varphi_3/\varphi_2$, the envelope duration and shape for a given input bandwidth can be described by $\varphi_2$ and $\alpha$.

To generate a narrowband SFG pulse with desired shape there are two main considerations when determining the sign and magnitude of the applied phases. First, to maintain the narrowband generation requirement during SFG, the two pulses must have equal and opposite amounts of $\varphi_2$ while the magnitude is given by solving
\begin{equation*}
    \Delta t = t_0\sqrt{1+\left( 4 \mathrm{ln} 2 \frac{\varphi_2}{t_0^2} \right)^2}
\end{equation*}
for $\varphi_2$ where $\Delta t$ is the final desired duration and t is the TL input duration. Second, the sign and magnitude of $\varphi_3$ are allowed to vary based on the desired SFG temporal envelope shape. Imparting the two copies with equal magnitude and opposite sign $\varphi_3$ results in symmetric profiles while allowing either sign or magnitude to vary results in asymmetric ones. While the absolute magnitudes of the imparted phases are dependent on input bandwidth and desired shaping, a larger $\alpha$ implies a more significant degree of shaping and a greater departure from the input pulse shapes.

\begin{figure}[ht!]
    \centering
    \includegraphics[width = 1\textwidth]{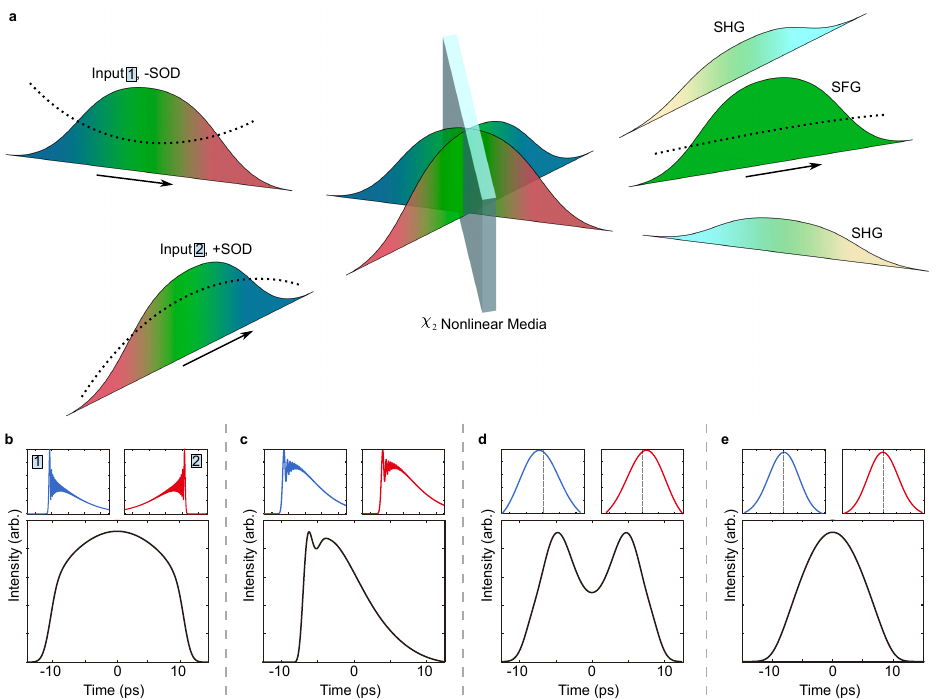}
    \caption{(a) Conceptual representation of DCNS where two oppositely chirped broadband pulses (phase represented with dashed lines) are incident on a $\chi_2$ nonlinear media. Two broadband chirped parasitic SHG pulses are generated along the same direction as the inputs and the narrowband near-TL SFG is generated at the mid-angle. (b-e) Four potential nonlinear shaping examples of the narrowband pulse via phase-only DCNS with varying SOD and TOD after a spectral bandpass filter of 0.5 nm. The temporal intensity profiles of input 1 and input 2 are shown in blue and red respectively. For more information on input pulse parameters and the applied spectral phase, see Supplementary Table 1.}
    \label{fig:Many_Shapes}
\end{figure}

In Figure 1 we showcase a selection of shaping possibilities with phase-only DCNS to exemplify the capabilities of this method. For each case, the shown intensity envelopes are after applying a 0.5 nm super-Gaussian spectral filter centered on the central frequency of the SFG to quench high-frequency oscillations inherent to the mixing process[SupMats]. The four scenarios that we present are (a) square-wave, (b) triangular, (c) double-hump, and (d) linear-ramp intensity distributions. Each scenario uses an input pulse with a different central wavelength and bandwidth to illustrate the applicability of DCNS to a wide range of laser systems. Of particular note are the triangular and double-hump cases for their direct applicability to high acceleration gradients in plasma wakefield acceleration\cite{tan2021formation} and emittance reduction in x-ray free electron laser facilities respectively\cite{penco2014experimental,dowell2016sources}. For the triangular case, the non-symmetric shape was achieved by having the same sign and magnitude of $\varphi_3$, and therefore opposite signed $\alpha$'s, on the two copies of the input. By having the same sign, the characteristic third-order temporal tails are aligned on the same side of the main peak. In the shown example, the pulse starts with a large intensity spike, potentially useful for driving optical shockwaves, but by flipping the sign of both inputs the temporal trace can be reversed in time to resemble a gradual ramp. However, any asymmetry generated this way also results in a residual phase on the SFG pulse potentially limiting the use of these pulses to non-phase-sensitive applications. In the double-hump scenario, the two inputs are functionally identical to the square wave with equal and opposite $\varphi_2$ and $\varphi_3$ but each has been shifted by 10 ps in time relative to the other to generate an intensity void. These examples are a small subset of potentially achievable temporal profiles given the large parameter ranges of $\varphi_2$, $\varphi_3$, time delay, and input pulse properties and serve to illustrate the flexibility behind DCNS rather than limit potential application.

\subsection{Experimental Phase-Only DCNS}


To experimentally verify DCNS, we generated a near-TL 20 ps flattop pulse in the UV from a commercial Yb:KGW laser system producing ~250 fs pulses centered at a wavelength of 1024 nm. Full details of this laser system can be found in Supplementary Note 1. To generate this temporal pulse profile, the required SOD and TOD applied to each pulse is $\pm$2.561 ps$^2$ and $\mp$0.28 ps$^3$, respectively, corresponding to an $\alpha = -.11$ ps. The desired temporal intensity profile of the SFG after a 0.5 nm bandpass filter is shown in Fig. \ref{fig:Many_Shapes} b. This shape is similar to a higher-order super-Gaussian pulse with a flat plateau across the central region of the pulse with rise and fall times of approximately 4 ps. These characteristics are highly sought after for reducing the emittance of photoemission-based free electron sources for ultrafast electron microscopy, high-energy physics, X-ray free electron lasers, and other secondary radiation sources.


\begin{figure}[ht!]
    \centering
    \includegraphics[width = 1\textwidth]{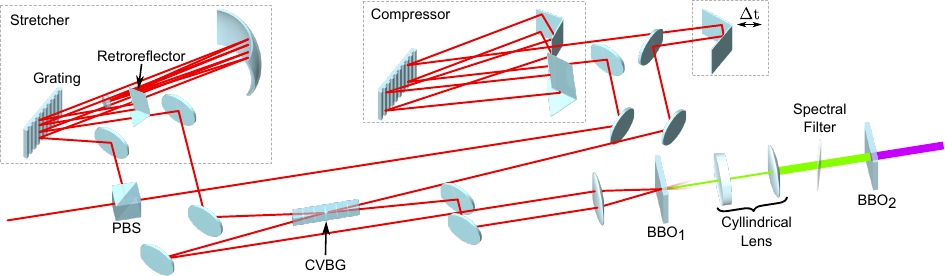}
    \caption{Experimental layout of phase shaping and nonlinear conversion elements used for generating a flattop in the UV. A single input beam is split with equal parts going through the grating stretcher and grating compressor. Each beam is then sent into opposite ends of the CVBG, reflected internally, and finally overlapped in BBO$_1$ for SFG. Temporal overlap in BBO$_1$ is achieved through a horizontal retroreflector on a linear delay stage($\Delta$t). The SFG beam travels through a cylindrical lens set and a spectral filter and is then converted to UV in BBO$_2$.}
    \label{fig:Experiment}
\end{figure}

Our experimental layout (Fig. \ref{fig:Experiment}) is divided into two main areas: spectral phase manipulation and nonlinear conversion. The phase manipulation portion is responsible for splitting the NIR input pulse into two distinct but equal copies, referred to as the stretcher and compressor pulses, and applying the equal and opposite spectral phases. For our input parameters, achieving an $\alpha$ $\sim$0.11 with this magnitude of SOD is a non-trivial task and unachievable within a single dispersive device utilizing gratings or prisms. We avoid this limitation by combining a matched grating compressor and stretcher designed to impart significantly more SOD than desired and the proper amount of TOD with a chirped volume Bragg grating (CVBG) designed to compensate for the residual SOD of both devices. After splitting the stretcher and compressor pulses are sent through the respective single grating devices and then reflected by opposite ends of the CVBG to realize the $\pm$2.561 ps$^2$ SOD and $\mp$0.28 ps$^3$ TOD. It is important to note that the layout we have chosen for our experimental demonstration is not the only viable layout for DCNS as a framework. More sophisticated methods for manipulating the individual phase of the input pulses could be conceptualized. Examples of such solutions are distinct synchronized laser systems or pulse arrays, individualized SLMs or AOMs for the inputs, or even mixing pulses from two separate amplification systems from the same master oscillator. Ultimately, manipulation of SOD, TOD, and spectral filtering on a per-pulse basis is all that is required for DCNS.

After phase manipulation the two pulses are sent through the nonlinear conversion chain in which the SFG pulse is generated, frequency-filtered, and further converted to the fourth harmonic of the NIR input. Each nonlinear conversion step is done in a critically phase-matched $\beta$-Barium borate (BBO) crystal. In the SFG step, the two pulses are made collinear and horizontally separated then focused by a single lens onto a 2 mm BBO at an angle of 1.5 degrees to each other. Time overlap is achieved with a horizontal retroreflecting delay stage in the compressor beam path. Though both NIR inputs are spatially Gaussian with no ellipticity, the SFG beam is generated with a significant ellipticity due to the crossing angle inside the crystal. After the residual NIR and parasitic SHG pulses are discarded, this ellipticity is corrected by passing the SFG pulse through a cylindrical lens before spectral filtering. The spectral filter is angle-tuned so that the center wavelength of the filter is the same as the SFG pulse. While a direct measurement of the SFG pulse's temporal intensity profile was not performed in this study, the narrowband characteristics and flat phase properties of the SFG pulse ensure the preservation of its temporal profile through subsequent nonlinear conversion stages. Therefore, the SFG pulse is converted to the fourth harmonic of the NIR via degenerate collinear SHG in a 3 mm BBO where the temporal intensity profile is measured.

\begin{figure}[ht!]
    \centering
    \includegraphics[width = 1\textwidth]{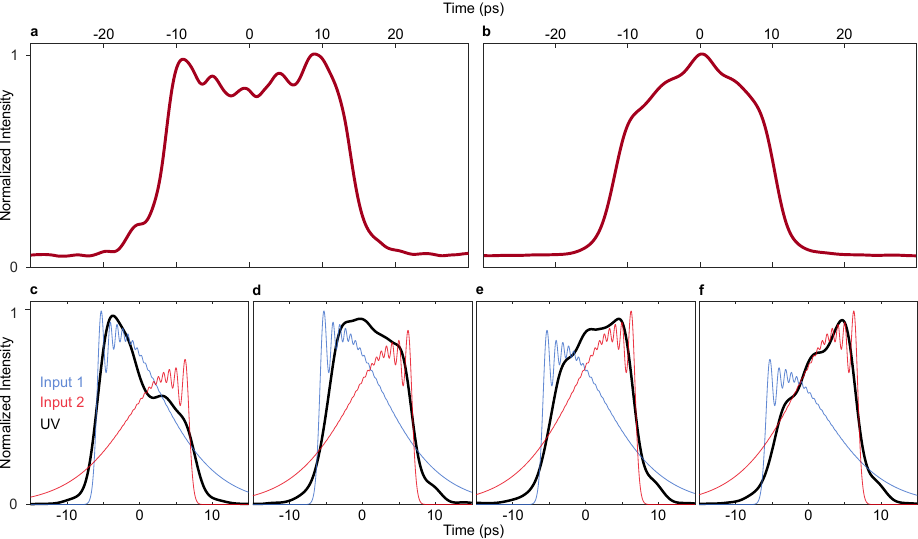}
    \caption{UV temporal intensity profile a) without IR filtering and b) with IR filtering. IR filtering is used to symmetrize the spectrum of both inputs post-interaction with the CVBG. (c-d) Once the spectrum is symmetrized power can be shifted along the UV pulse (black) by adjusting the relative amplitudes of IR inputs 1 (blue) and 2 (red). IR input profiles are simulated from the expected spectral phase after the stretcher/compressor and CVBG.}
    \label{fig:UV_Shape}
\end{figure}

The UV temporal profile is measured via third-harmonic cross-correlation with a 70 fs, 1035 nm pulse from the oscillator feeding the main laser amplifier. In Figure \ref{fig:UV_Shape}a, the initial temporal profile of the UV from the above setup is shown. While this profile does demonstrate the sharp rise and fall times that were seen in the simulation, the flat central region exhibits noticeable intensity modulations. These modulations originated from an imperfection in the CVBG that removed a small amount of the long-wavelength side of the stretcher and compressor pulses. This loss of information resulted in an unbalanced contribution of the short wavelengths and thus higher frequency oscillations across the temporal profile. To correct this, we inserted a long-pass filter before the SFG focusing lens to remove a small amount of the short-wavelength side of the NIR inputs. Once the spectrum of the inputs was symmetrized, the resulting UV profile (Fig. \ref{fig:UV_Shape}b) drastically smoothed across the central region while maintaining the sharp rise and fall times as desired. Additionally, peak intensity can be shifted in time throughout the pulse by only varying the relative energy of the two input pulses with respect to one another. From the baseline flattop shape (Fig. \ref{fig:UV_Shape}b), the rising and falling edge can be preferentially enhanced in a continuous fashion as shown in the series Fig. \ref{fig:UV_Shape}.



\subsection{Experimental Phase and Amplitude DCNS Shaping}

\begin{figure}[ht!]
    \centering
    \includegraphics[width = 0.9\textwidth]{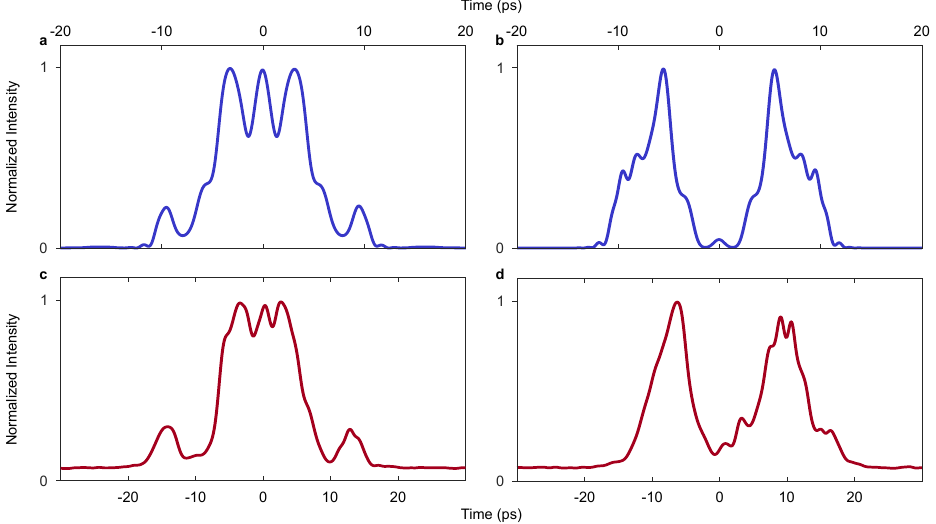}
    \caption{Simulation of direct temporal intensity shaping during DCNS due to spectral amplitude shaping of an input pulse with a) two spectral holes, one on either side of the central frequency and b) a single spectral hole at the central frequency. Experimental spectral amplitude shaping for the c) two spectral holes, one on either side of the central frequency and d) a single spectral hole at the central frequency.}
    \label{fig:Amp_Shaping}
\end{figure}

Affecting the UV pulse shape in this way revealed an important extension of the DCNS process towards arbitrary pulse shaping: programmable spectral amplitude filtering. In Fig. 4, we illustrate the potential of applying amplitude masking on top of a DCNS-shaped pulse. In this example, two scenarios of spectral holes being applied to the pulse with negative SOD before SFG were investigated: two spectral holes each $±1.5$ nm away from the central wavelength with widths of 0.5 nm (Fig. \ref{fig:Amp_Shaping} left), and a single spectral hole with a 1 nm width at the central wavelength of the pulse (Fig. \ref{fig:Amp_Shaping} right). Figure \ref{fig:Amp_Shaping}(a,b) are simulated results of applying these two filters to the flattop baseline DCNS pulse, and Fig. \ref{fig:Amp_Shaping}(c,d) are experimental demonstrations of the same spectral filters. In the case of the double hole, the envelope develops a strong intensity modulation across the central peak flanked by two small pre- and post-pulses. For the single spectral hole, the SFG pulse has a temporal profile with a deep central valley and sharply rising triangular side lobes. Experimentally, both cases are well matched in qualitative shape with the double hole generating the tri-modulated central peak with flanking sub-pulses, and the single spectral hole generating the double-peaked pulse. While the experimentally generated pulses have the same qualitative shape they are slightly longer than the simulated results. This deviation likely derives from a greater SOD and $\alpha$ from the matched grating pair than the SOD and $\alpha$ in simulation.

For both of these cases, intensity is reduced but does not go to zero at temporal locations corresponding to the location of the holes within the spectrum. While the SOD applied to these pulses is significant, especially when discussing ultrashort optics, each frequency component still temporally overlaps with those around it by some amount. As such, there is still some power within the temporal region from frequency components that haven't been filtered out. Should the magnitude of applied SOD grow, the mapping of spectrum-to-time becomes closer to one-to-one and the shaping more precise. However, this effect is muted by the spatial overlap and lack of perfect spectral filters in the Fourier plane of the compressor grating where the spectral filtering is applied. Lastly, like the baseline flattop scenario, these temporal traces are collected from in the UV rather than directly after SFG indicating that even with spectral shaping, the SFG pulse maintains its narrow-band, flat-phase nature. An intrinsic limitation of spectral amplitude shaping is decreasing the input pulse energy to such a degree that efficient nonlinear conversion in the SFG stage is no longer possible. However, by combining this with phase-only DCNS, the degree of potential shaping is extended beyond what can be achieved in shaping methods relying solely on amplitude masking for high energy or high average power applications.


\section{Discussion}

We have described and demonstrated DCNS a new methodology that can generate pulses with tailored temporal envelope shapes in the picosecond range. By utilizing non-collinear SFG with oppositely chirped inputs, we avoid the pitfalls of previous methods, such as residual spectral chirp, phase discontinuities, or unwanted intensity fluctuations, while integrating the spectral phase and amplitude shaping to enable flexibility of achievable temporal intensity profiles. As a proof-of-principle experiment based on this framework, we generated a 20 ps temporally-flattop UV pulse with little intensity fluctuations and highly compressed spectral bandwidth from 240 fs 1024 nm inputs. Additionally, we demonstrated a potential avenue toward adaptive temporal shaping through spectral amplitude modification of the broadband inputs before non-linear conversion. Our experimental implementation of DCNS is only one of many potential solutions with the most immediate improvements being found by eliminating the requirement of a CVBG for correcting large amounts of SOD. By circumventing the limitations of prior attempts at bridging the picosecond shaping gap, our method is readily deployable for a plethora of applications such as patterning of picosecond duration pulses driving nanoplasma excitation, enhancement of ultrafast photoelectron sources such as XFELs, and laser-driven artificial photosynthesis.


\section{Methods}

\subsection{Nonlinear Conversion Modeling}

The dynamics involved with DCNS include laser propagation, dispersion, and nonlinear conversion in millimeter-scale nonlinear crystals and are described by the one-dimensional nonlinear Schr\"odinger equation~\cite{NLSEarxiv,agrawal2000nonlinear} (1D-NLSE). To model this we utilized the symmetrized split-step Fourier method (S-SSFM)\cite{weideman1986split} which oscillates between the Fourier conjugate domains of time and frequency for calculating nonlinear dynamics and dispersion respectively as the numerical method propagates along the crystal. Propagation along the crystal is handled via the angular spectrum method in the spatial frequency domain. Applying the S-SSFM to the 1D-NLSE, the governing equation for our dynamics is given by 
\begin{equation}
    A(z+dz,t) = \mathcal{F}^{-1}\left[e^{i \frac{C_{D}}{2} \omega^2 dz} \mathcal{F}\left[e^{iC_{NL}|A|^2dz} \mathcal{F}^{-1}\left[ e^{i \frac{C_{D}}{2} \omega^2 dz} \mathcal{F}\left[ A(z,t) \right]\right]\right]\right]\label{eqn:PropSym}
\end{equation}
where $\mathcal{F}$ and$\mathcal{F}^{-1}$ represent the forward and reverse Fourier transform respectively, $A$ describes the pulse envelope in time, $t$, and space, $z$, $C_{D}$ is the dispersion coefficient and governs the linear portion of the equation, and $C_{NL}$ is the nonlinear coefficient. While the linear and nonlinear portions and their effects are not separable \textit{a priori}, the error generated in handling their effects separately is small as long as the numerical step size, $dz$, is sufficiently small\cite{agrawal2000nonlinear}.

Each DCNS scenario was modeled as the interaction between three fields, the two dispersed input fields and the SFG field with each being described individually by their respective 1D-NLSE equations. The nonlinear conversion and inter-field interactions was handled by the coupled equations for SFG\cite{boyd2019nonlinear},
\begin{subequations}
    \begin{align}
        \frac{d A_1}{d z} &= \frac{2 i d_{eff} \omega_1^2}{k_1 c^2}A_2^* A_3 e^{-i \Delta k z} \\
        \frac{d A_2}{d z} &= \frac{2 i d_{eff} \omega_2^2}{k_2 c^2}A_1^* A_3 e^{-i \Delta k z} \\
        \frac{d A_3}{d z} &= \frac{2 i d_{eff} \omega_3^2}{k_3 c^2}A_1 A_2 e^{i \Delta k z}
    \end{align}
    \label{eqn:SFGEqns}
\end{subequations}
where $d_{eff}$ is the effective nonlinearity of the crystal media and $\Delta k$ is the phase-mismatch between the involved fields. Our implementation of the S-SSFM calculated the full envelope of the three fields in the time and frequency domains allowing for explicitly tracking the phase mismatch between each spectral component of each field with respect to the full spectrum of the other two. As such, the monochromatic plane-wave assumptions underlying the coupled SFG equations are respected.

\section{Supplemental Materials}

\subsection{Supplementary Table 1}
\begin{table}[ht]
\begin{tabular}{|c|c|c|c|c|c|}
\hline
            & \begin{tabular}[c]{@{}c@{}}Input Central\\ Wavelength (nm)\end{tabular} & \begin{tabular}[c]{@{}c@{}}Input Pulse\\ Duration (fs)\end{tabular} & \begin{tabular}[c]{@{}c@{}}Pulse 1 SOD ($ps^2$)\\ and $\alpha$ (ps)\end{tabular} & \begin{tabular}[c]{@{}c@{}}Pulse 2 SOD ($ps^2$)\\ and $\alpha$ (ps)\end{tabular} & Time Delay (ps) \\ \hline
Flattop     & 1024                                                                    & 246                                                                 & -2.561, -0.11                                                                    & 2.561, -0.11                                                                     & 0               \\ \hline
Triangular  & 800                                                                     & 400                                                                 & 2.1, 0.16                                                                        & -2.1, -0.16                                                                      & 0               \\ \hline
Double Hump & 1550                                                                    & 70                                                                  & 0.75, -0.0036                                                                    & -0.75, -0.0036                                                                   & 1               \\ \hline
Linear Ramp & 1064                                                                    & 40                                                                  & 0.2, -0.001                                                                      & -0.2, -0.001                                                                     & 0               \\ \hline
\end{tabular}
\end{table}

\subsection{Supplementary Note 1}
The laser system is a 40 W Light Conversion Carbide. This system can maintain 40 W of average power at repetition rates between 100 kHz and 1 MHz adjusting the pulse energy between 400 uJ and 40 uJ as required. The central wavelength is ~1024 nm with a bandwidth of 8 nm at 40 uJ and 7 nm at 400 uJ.

\subsection{Supplementary Note 2}
Due to the large intensity fluctuations of the input pulses when $\alpha$ is high, and therefore shaping contributions from TOD is significant, the sum-frequency generated pulse also has intrinsic fast intensity oscillations. These oscillations are due to asymmetrical high-frequency spectral content on the SFG pulse. By applying a low pass filter (spectral filter) in frequency space the main temporal profile of the pulse can be retrieved while damping intensity oscillations from the high-frequency tail.

\section{Acknowledgements}
We thank Michael Greenberg for his help in designing the layout of the folded stretcher and compressor, Shawn Alverson for his contributions to automation and controls integration, and Sasha Gilevich for her guidance in the setup and alignment of the cross-correlator. This work was funded by the Department of Energy Basic Energy Sciences under contract numbers DE-AC02-76SF00515 and DE-SC0022559,  the Air Force Office of Scientific Research under contract no. FA9550-23-1-0409 and the Office of Naval Research under contract no. N00014-24-1-2038. Jack Hirschman would like to acknowledge support under the DoD NDSEG Fellowship.

\section{References}

\bibliography{bib}
\bibliographystyle{unsrt}

\end{document}